\documentclass[aps,amsmath,amssymb, prc,twocolumn,superscriptaddress]{revtex4-1}

\usepackage{graphicx}
\usepackage{dcolumn}
\usepackage{bm}
\usepackage[utf8]{inputenc}
\usepackage[english]{babel}
\usepackage{float}
\usepackage{multirow}
\usepackage{longtable}
\usepackage{dcolumn}
\newcolumntype{d}{D{.}{.}{-1}}
\usepackage{setspace}
\usepackage{threeparttable}
\usepackage{tabularx}
\usepackage[symbol*]{footmisc}
\DefineFNsymbolsTM{myfnsymbols}{
  \textasteriskcentered *
  \textdagger    \dagger
  \textdaggerdbl \ddagger
  \textsection   \mathsection
  \textbardbl    \|%
  \textparagraph \mathparagraph
}%
\setfnsymbol{myfnsymbols}

\usepackage{siunitx}
\newcommand{\nuc}[2]{\hbox{$^{#1}$#2}}

\usepackage{gensymb}

\begin{document}


\title{Investigation of octupole collectivity near the $A =72$ shape-transitional point}


\author{M. Spieker}
\email[]{mspieker@fsu.edu}
\affiliation{Department of Physics, Florida State University, Tallahassee, Florida 32306, USA}

\author{L.A. Riley}
\affiliation{Department of Physics and Astronomy, Ursinus College, Collegeville, PA 19426, USA}

\author{P.D. Cottle}
\affiliation{Department of Physics, Florida State University, Tallahassee, Florida 32306, USA}

\author{K.W. Kemper}
\affiliation{Department of Physics, Florida State University, Tallahassee, Florida 32306, USA}

\author{D. Bazin}
\affiliation{Facility for Rare Isotope Beams, Michigan State University, East Lansing, Michigan 48824, USA}
\affiliation{Department of Physics and Astronomy, Michigan State University, East Lansing, Michigan 48824, USA}

\author{S. Biswas}
\affiliation{Facility for Rare Isotope Beams, Michigan State University, East Lansing, Michigan 48824, USA}



\author{P.J. Farris}
\affiliation{Facility for Rare Isotope Beams, Michigan State University, East Lansing, Michigan 48824, USA}
\affiliation{Department of Physics and Astronomy, Michigan State University, East Lansing, Michigan 48824, USA}

\author{A. Gade}
\affiliation{Facility for Rare Isotope Beams, Michigan State University, East Lansing, Michigan 48824, USA}
\affiliation{Department of Physics and Astronomy, Michigan State University, East Lansing, Michigan 48824, USA}

\author{T. Ginter}
\affiliation{Facility for Rare Isotope Beams, Michigan State University, East Lansing, Michigan 48824, USA}

\author{S. Giraud}
\affiliation{Facility for Rare Isotope Beams, Michigan State University, East Lansing, Michigan 48824, USA}

\author{J. Li}
\affiliation{Facility for Rare Isotope Beams, Michigan State University, East Lansing, Michigan 48824, USA}

\author{S. Noji}
\affiliation{Facility for Rare Isotope Beams, Michigan State University, East Lansing, Michigan 48824, USA}

\author{J. Pereira}
\affiliation{Facility for Rare Isotope Beams, Michigan State University, East Lansing, Michigan 48824, USA}

\author{M. Smith}
\affiliation{Facility for Rare Isotope Beams, Michigan State University, East Lansing, Michigan 48824, USA}

\author{D. Weisshaar}
\affiliation{Facility for Rare Isotope Beams, Michigan State University, East Lansing, Michigan 48824, USA}

\author{R.G.T. Zegers}
\affiliation{Facility for Rare Isotope Beams, Michigan State University, East Lansing, Michigan 48824, USA}
\affiliation{Department of Physics and Astronomy, Michigan State University, East Lansing, Michigan 48824, USA}


\date{\today}

\begin{abstract}

Enhanced octupole collectivity is expected in the neutron-deficient Ge, Se and Kr isotopes with neutron number $N \approx 40$ and has indeed been observed for \nuc{70,72}{Ge}. Shape coexistence and configuration mixing are, however, a notorious challenge for theoretical models trying to reliably predict octupole collectivity in this mass region, which is known to feature rapid shape changes with changing nucleon number and spin of the system. To further investigate the microscopic configurations causing the prolate-oblate-triaxial shape transition at $A \approx 72$ and their influence on octupole collectivity, the rare isotopes \nuc{72}{Se} and \nuc{74,76}{Kr} were studied via inelastic proton scattering in inverse kinematics. While significantly enhanced octupole strength of $\sim 32$ Weisskopf units (W.u.) was observed for \nuc{72}{Se}, only strengths of $\sim 15$ W.u. were observed for \nuc{74,76}{Kr}. In combination with existing data, the new data clearly question a simple origin of enhanced octupole strengths around $N = 40$. The present work establishes two regions of distinct octupole strengths with a sudden strength increase around the $A=72$ shape transitional point.

\end{abstract}


\maketitle



\section{Introduction}

Much of the study of the structure of atomic nuclei centers on the interplay between individual nucleons and the emergent collective behavior caused by the strong interaction between them. Quadrupole-deformed shapes are one of the emergent phenomena. Among these, axially-symmetric prolate (cigar-like) shapes are observed more frequently than oblate (disk-like) shapes \cite{Ham09a, Hor10a}. In addition, axially-asymmetric (triaxial) shapes are important in some regions of the nuclear chart, including the Ge-Kr mass region ($Z = 32-36$) \cite{Lec78a, Lec80b, Kot90a, Guo07a, Gir09a, Rod14a, Nik14a, Aya16a, Hen19a, Aya19a}. In some of the nuclei in this region, both axially symmetric and asymmetric shapes appear to coexist at comparably low excitation energies and lead to complex quantum-state mixing \cite{Hey11a, Goe16a, Gar21a}. The delicate interplay between the different configurations influences several experimental observables connected to the quadrupole degree of freedom and, furthermore, causes rapid shape changes observed with both isospin and spin \cite{Gad05a, Cle07a, Lju08a, Iwa14a, Hen18a, Wim20a, Wim21a}. 

In addition to quadrupole excitations, octupole excitations are observed throughout the nuclear chart \cite{But96a, Rob11a, But16a, Cao20a}. Due to the presence of the $2p_{3/2}$ and $1g_{9/2}$ orbitals for both protons and neutrons around the Fermi surface, enhanced electric octupole $B(E3)$ transition strengths are expected for the neutron-deficient Ge, Se and Kr isotopes. Previous experimental studies established that the low energy octupole state (LEOS) fragments into two or more $J^{\pi} = 3^-_i$ states with the $B(E3;3^-_i \rightarrow 0^+_1)$ strengths summing up to approximately 15 Weisskopf units (W.u.) \cite{Kib02a, Lec80a, Mat82a, Ros86a, Ogi86a, Sch87a}. However, \nuc{70,72}{Ge} are notable exceptions as a very sudden $B(E3; 3^-_1 \rightarrow 0^+_1)$ strength increase to around 30 W.u. is observed \cite{Lec80a, Ros86a, Sch87a}. Chuu {\it et al.} were able to describe this $B(E3)$ strength increase in the Ge isotopes with the Interacting-Boson-Model plus Interacting-Boson-Fermion-Model approach (IBM+IBFM); however, without considering shape coexistence \cite{Chu93a}. In their study, they attributed the sudden increase in octupole collectivity to a maximum contribution of the collective $f$-boson configuration to the total wave function. The contribution of the $f_{5/2}-g_{9/2}$ fermion-pair configuration turned out to be negligibly small at $N=40$. Interestingly though, the sudden strength increase is not observed for the $N=40$ isotone \nuc{74}{Se} \cite{Ogi86a}. When comparing to the other isotonic chains, there also appears to be nothing particularly special about proton number $Z = 32$ in terms of octupole collectivity \cite{Kib02a}. This questions previous conclusions about a simple origin of enhanced octupole collectivity at $N=40$ drawn in, {\it e.g.}, Refs. \cite{Chu93a, Wan22a, Cot90c}. Instead, the idea that octupole collectivity is more sensitive to quadrupole distortions in the Ge-Kr region than in other mass regions might be correct \cite{Naz90a}. Up to now, the sharp difference in octupole collectivity between \nuc{70,72}{Ge} and the rest of the nuclei in this region has remained a puzzle. Shape coexistence and strong configuration mixing generally complicate the theoretical description of octupole strengths (see, {\it e.g.}, the remarks in \cite{Rob12a}). To more systematically approach this challenge, first exploratory calculations within the framework of the configuration-mixing $sdf$ IBM mapping approach, which is based on microscopic self-consistent mean-field calculations employing universal energy density functionals and takes shape coexistence explicitly into account, have recently been performed \cite{Nom22b}. In addition to enhanced strength for \nuc{72}{Ge}, $B(E3; 3^-_1 \rightarrow 0^+_1)$ strengths of around 30 W.u. have been predicted in the rare isotopes \nuc{74,76}{Kr}.

In this work, we report on measurements of the previously unknown $B(E3;3^-_i \rightarrow 0^+_1)$ strengths in the rare isotopes \nuc{72}{Se} ($Z = 34$) and \nuc{74,76}{Kr} ($Z = 36$), which are the $N=38$ and $N=40$ isotones of \nuc{70,72}{Ge}. To measure the octupole strengths in these nuclei, inelastic proton scattering experiments in inverse kinematics were performed. Inelastic proton scattering has proven to be a very powerful tool to study the fragmentation of the LEOS among a few to several excited 3$^-$ states for different structures of the ground state \cite{Ogi86a, Cot88a, Cot90a, Ken92a}. In combination with data available for stable nuclides, the new data clearly show that a simple picture of enhanced octupole correlations around the octupole magic number $N = 40$ cannot be claimed.

\section{Experiment}

The experiments were performed at the Coupled Cyclotron Facility of the National Superconducting Cyclotron Laboratory (NSCL) at Michigan State University\,\cite{NSCL} at secondary beam energies corresponding to proton energies of around 100 MeV in the center-of-mass frame. At these energies, both proton and neutron contributions to the wave function are probed almost equally \cite{Cot01a}.  The secondary \nuc{76}{Kr} (79\,$\%$ purity), \nuc{74}{Kr} (51\,$\%$ purity) and \nuc{72}{Se} (6\,$\%$ purity) beams were produced from a 150\,MeV/u $^{78}$Kr primary beam in projectile fragmentation on a 308-mg/cm$^2$ thick $^{9}$Be target. The A1900 fragment separator\,\cite{Mor03}, using a 240-mg/cm$^2$ Al degrader, was tuned to select the fragments of interest in flight using two separate magnetic settings. For the magnetic setting centered on \nuc{74}{Kr}, the secondary \nuc{72}{Se} beam was part of the cocktail beam. All three secondary beams could be unambiguously distinguished from the other components in the cocktail beam via the time-of-flight difference measured between two plastic scintillators located at the exit of the A1900 and the object position of the S800 analysis beam line. Downstream, the NSCL/Ursinus Liquid Hydrogen (LH$_2$) Target was located at the target position of the S800 spectrograph. The projectilelike reaction residues entering the S800 focal plane were identified event-by-event from their energy loss and time of flight\,\cite{Baz03}. 

\begin{figure}[t]
\centering
\includegraphics[width=1\linewidth]{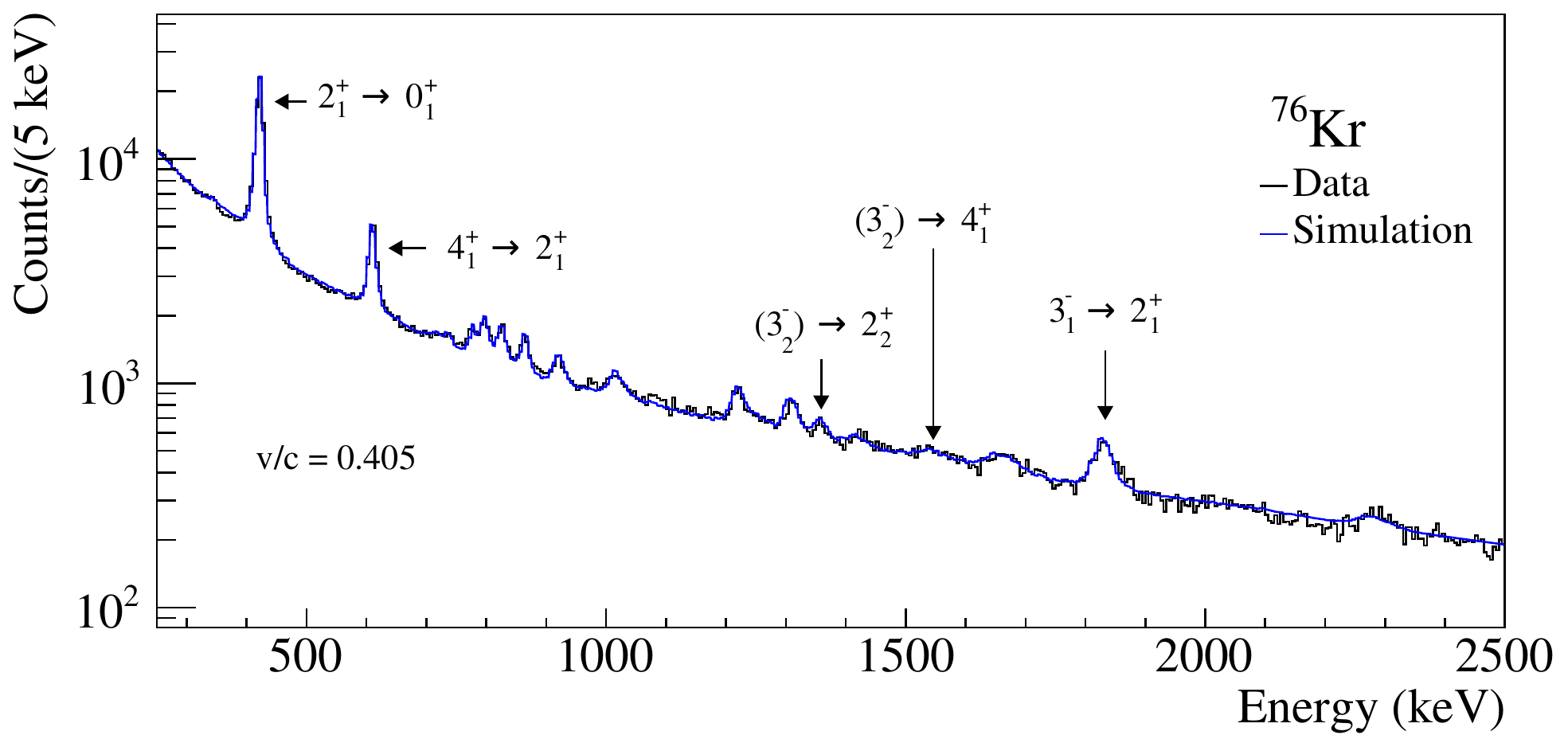}
\includegraphics[width=1\linewidth]{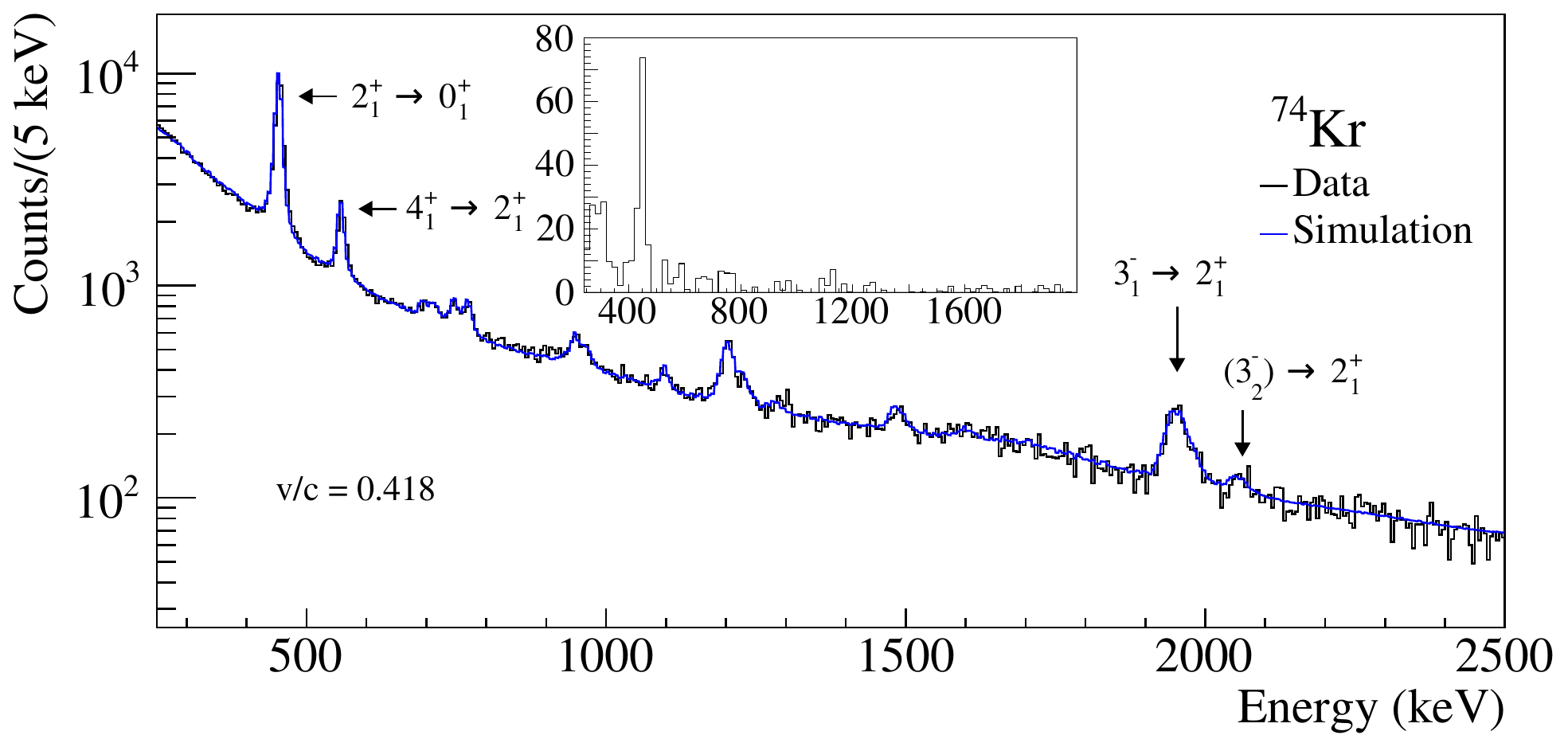}
\includegraphics[width=1\linewidth]{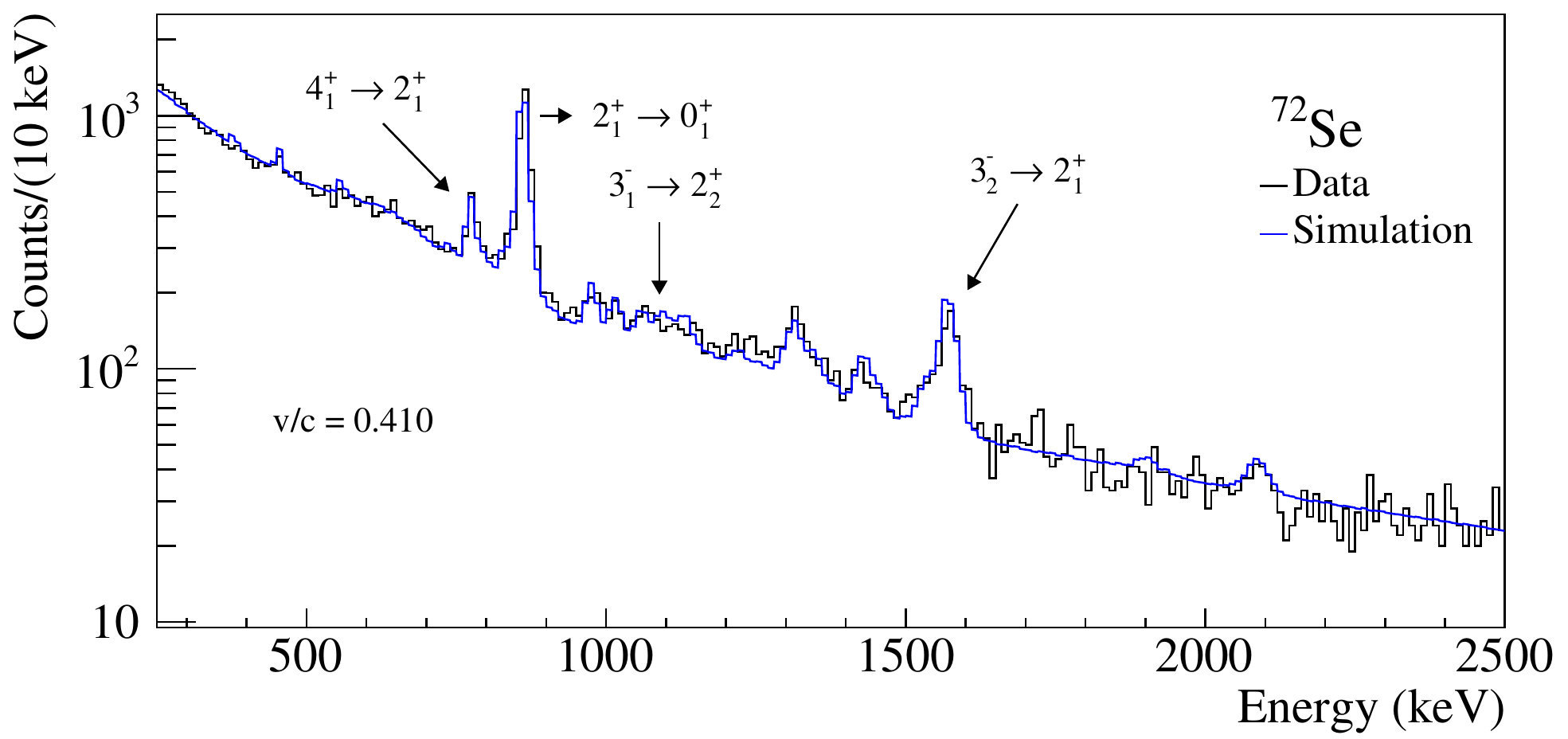}
\caption{\label{fig:spectra}{(color online) Doppler-corrected, in-beam $\gamma$-ray spectra for \nuc{76}{Kr} (top), \nuc{74}{Kr} (middle) and \nuc{72}{Se} (bottom). Data are shown in black. \textsc{geant4} simulations performed with \textsc{ucgretina} \cite{Ril21a} are presented in blue. A prompt background consisting of two exponential functions was included in the simulation. As an example for the $\gamma \gamma$ coincidences placing the $3^-_1$ state in \nuc{74}{Kr}, the inset in the middle panel shows the coincidence spectrum when gated on the 1953-keV, $3^-_1 \rightarrow 2^+_1$ $\gamma$-ray transition. A clear coincidence with the 456-keV, $2^+_1 \rightarrow 0^+_1$ $\gamma$-ray transition is observed.}} 
 \end{figure}

The GRETINA $\gamma$-ray tracking array \cite{Pas13a,Wei17a} was used to detect $\gamma$ rays emitted by the reaction residues in flight ($v/c \approx 0.4$). Eight GRETINA modules, containing four, 36-fold segmented HPGe detectors each, were mounted in the north half of the mounting shell to accommodate the LH$_2$ target. Event-by-event Doppler reconstruction of the residues' $\gamma$-ray energies was performed based on the angle of the $\gamma$-ray emission determined from the main interaction point in the Ge crystal and including trajectory reconstruction of the residues through the S800 spectrograph \cite{Wei17a}. Fig.\,\ref{fig:spectra} shows the experimental Doppler-corrected, in-beam $\gamma$-ray spectra for \nuc{72}{Se} and \nuc{74,76}{Kr} together with the corresponding spectra simulated with \textsc{ucgretina} \cite{Ril21a}. For the simulation, the known experimental kinematics, target thickness, setup geometry and $\gamma$-ray detection efficiency were used as inputs. Pressure differences across the Kapton entrance and exit windows of the LH$_2$ cell cause them to bulge outwards. To determine the target thickness, this effect was taken into account and its contribution quantified by simulating the kinetic-energy distribution of the outgoing beam with the procedure described in \cite{Ril19a}. An areal density of 69(3) mg/cm$^2$ was determined. Assuming the population of different excited states in the reaction, the experimental $\gamma$-ray yields were determined by fitting a superposition of the simulated $\gamma$-decay spectra of individual excited states to the experimental spectrum. For each excited state, $\gamma$-decay branching was explicitly taken into account if known from previous experiments \cite{ENSDF, Gia05a, McC11a, Dun13a, Muk22a}. The $\gamma$-decay intensities were varied within the reported uncertainties.  As $\gamma$-ray cascades are included in the simulation, the obtained yields are corrected for observed feeders. The yields are used to calculate the inelastic proton scattering cross sections to excited states in \nuc{72}{Se} and \nuc{74,76}{Kr} by normalizing them to the number of incoming beam particles and the number of target nuclei. For the $2^+_1$ states of \nuc{72}{Se} and \nuc{74,76}{Kr}, the $(p,p')$ cross sections are 17(4)\,mb, 28(5)\,mb and 43(2)\,mb, respectively. For the $3^-_1$ states of \nuc{74,76}{Kr} and the $3^-_2$ state of \nuc{72}{Se}, they are 4.6(8)\,mb, 5.9(3)\,mb and 13(3)\,mb. Stated uncertainties include statistical uncertainties, the stability of the secondary beam composition, uncertainties coming from the choice of software gates and the target thickness.

\section{Results and Discussion} 

\begin{figure}[t]
\centering
\includegraphics[width=1\linewidth]{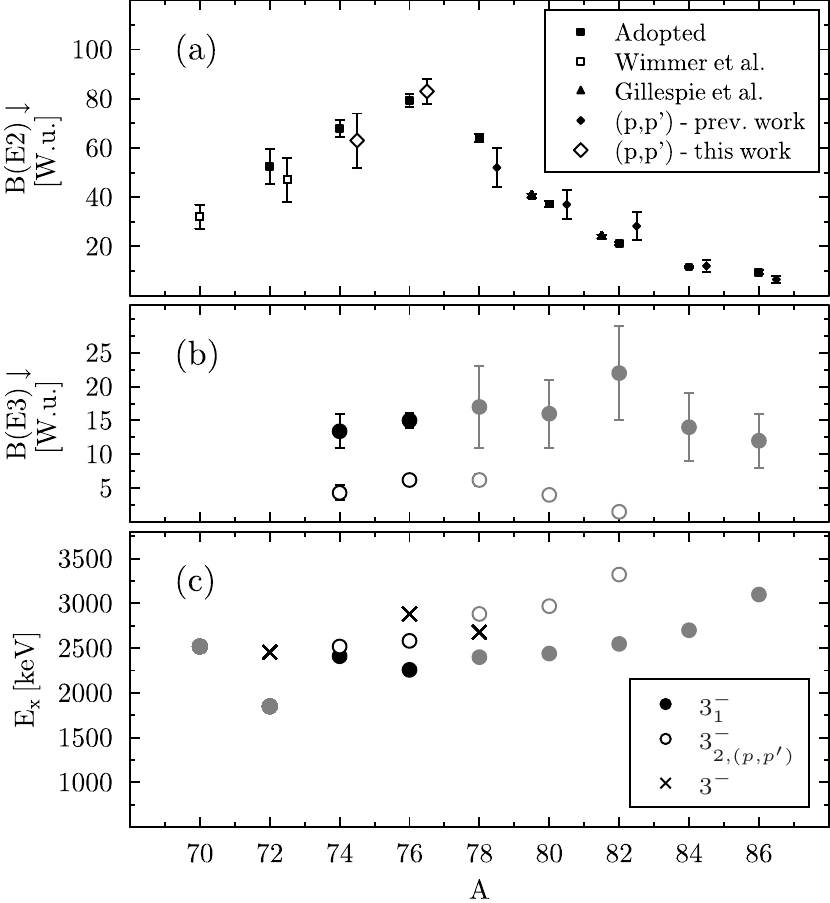}
\caption{\label{fig:systematics}{Experimental systematics of (a) $\mathrm{B(E2;2^+_1 \rightarrow 0^+_1)}$, (b) $\mathrm{B(E3;3^-_i \rightarrow 0^+_1)}$ strengths and (c) excitation energies $\mathrm{E_x}$ for the $3^-_1$ states and $3^-_{2,(p,p')}$ states in Kr isotopes. In panels (b) and (c), data obtained in this work are shown in black and previously reported data in grey \cite{Mat82a}. The notation $3^-_{2,(p,p')}$ is used to indicate that they are the second-strongest fragment observed in $(p,p')$. The $3^-_{2,(p,p')}$ states in \nuc{74,76}{Kr}, observed in this work, are tentatively assigned as discussed in the text. Additional $3^-$ states \cite{ENSDF} important for the discussion are presented with a cross in panel (c). As seen in panel (c), the $3^-_{2,(p,p')}$ states do not necessarily correspond to the second $3^-$ state. Other $B(E2)$ data are from \cite{Pri14a, Wim20a, Gil21a, Wim21a, Sak79a}.}} 
 \end{figure}

To calculate reduced transition probabilities $B(E\lambda)$ from deformation parameters $\beta_{\lambda}$ ($\lambda = 2,3,4,...$), reaction calculations were performed with the coupled-channels program \textsc{chuck3} \cite{chuck} using the global optical-model parameters of \cite{Kon03a}. Only single-step excitation was considered. As described in Refs. \cite{Ken92a, Kib02a, Pri14a}, the deformation parameters $\beta_{\lambda}$ can be calculated by scaling the theoretical cross sections to the experimentally determined ones. For the $2^+_1$ states, deformation parameters of $\beta_2 = 0.40 \pm 0.02 \ \mathrm{(stat.)} \pm 0.03 \ \mathrm{(sys.)}$ for \nuc{76}{Kr}, $\beta_2 = 0.35 \pm 0.06 \ \mathrm{(stat.)} \pm 0.02 \ \mathrm{(sys.)}$ for \nuc{74}{Kr}, and $\beta_2 = 0.26 \pm 0.06 \ \mathrm{(stat.)} \pm 0.02 \ \mathrm{(sys.)}$ for \nuc{72}{Se} were determined. Systematic uncertainties stem from the theoretically expected variation of the cross section over the LH$_2$ target thickness. Within uncertainties, the $\beta_2$ values for the Kr isotopes are in excellent agreement with the adopted values of 0.3920(66) and 0.363(9) \cite{Pri14a}, respectively. For \nuc{72}{Se}, the value agrees with the adopted value of 0.215(5) within uncertainties. For \nuc{74,76}{Kr}, the inferred $B(E2;2^+_1 \rightarrow 0^+_1)$ values are shown in Fig.\,\ref{fig:systematics}\,(a) including data for the Kr isotopes between $N=34$ and $N=50$. As can be seen, the $(p,p')$ data confirm the trend of a decreasing $B(E2)$ strength when passing $N=40$ ($A = 76$) and agree with the strengths determined with other probes \cite{Pri14a, Wim20a, Gil21a, Wim21a, Sak79a}, validating both the reaction calculations and feeding correction.

The 2257-keV, $3^-_1$ state of \nuc{76}{Kr} was previously observed in several experiments including the $(p,t)$ experiment of \cite{Mat81a}. No excited $3^-$ states were known in \nuc{74}{Kr} prior to this work. The first two excited $3^-$ states of \nuc{72}{Se} at 2406\,keV and 2434\,keV, respectively, were observed in a number of experiments \cite{ENSDF, Abr10a}. However, for none of the three nuclei, $B(E3)$ strengths were previously measured. In this work, the well-known 1834-keV, $3^-_1 \rightarrow 2^+_1$ $\gamma$-ray transition in \nuc{76}{Kr} is prominently observed (see Fig.\,\ref{fig:spectra}). The $J^{\pi} = 3^-_1$ state of \nuc{74}{Kr} is newly assigned based on the striking similarity of its 1953-keV, $3^-_1 \rightarrow 2^+_1$ $\gamma$-ray transition to the corresponding one in \nuc{76}{Kr} (also see Fig.\,\ref{fig:spectra}). $\gamma \gamma$ coincidences confirm the placement and establish the \nuc{74}{Kr}, $J^{\pi} = 3^-_1$ state at 2409(3)\,keV (see inset in Fig.\,\ref{fig:spectra}). Reduced transition strengths $B(E3;3^-_1 \rightarrow 0^+_1)$ of $15.0 \pm 0.9 \mathrm{(stat.)} \pm 1.8 \mathrm{(sys.)}$ W.u. for \nuc{76}{Kr} and $13 \pm 2 \mathrm{(stat.)} \pm 2 \mathrm{(sys.)}$ W.u. for \nuc{74}{Kr} were determined, respectively. As can be seen in Fig.\,\ref{fig:systematics}\,(b), they match the rather constant $B(E3;3^-_1 \rightarrow 0^+_1)$ values of around 15 W.u. observed in the stable Kr isotopes \cite{Mat82a}. The excitation-energy systematics in the Kr isotopes are presented in Fig.\,\ref{fig:systematics}\,(c). Interestingly, rather than the $3^-_1$ state, the $3^-_2$ state is the most strongly populated $3^-$ state in \nuc{72}{Se} (see 1572-keV, $3^-_2 \rightarrow 2^+_1$ $\gamma$-ray transition in the bottom panel of Fig.\,\ref{fig:spectra}). A significantly larger $B(E3;3^-_2 \rightarrow 0^+_1)$ strength of $32 \pm 7 \mathrm{(stat.)} \pm 4 \mathrm{(sys.)}$ W.u. is determined. For the $3^-_1$ state, only an upper limit of $B(E3;3^-_1 \rightarrow 0^+_1) \leq 4.5$ W.u. can be reported. In agreement with newer experiments on \nuc{72}{Se} \cite{McC11a, Muk22a}, the $3^-_2 \rightarrow 0^+_1$ ground-state branch is not observed. Note that in \nuc{74,76}{Kr}, the $3^-_1 \rightarrow 0^+_1$ $\gamma$-decay branch was also not observed and, thus, not considered for the calculation of the experimental $(p,p')$ cross sections.

Given that the $3^-_2$ state is more strongly populated than the $3^-_1$ state in \nuc{72}{Se}, it is worth noting that Ref. \cite{Mat82a} established a pronounced variation of the $B(E3;3^-_2 \rightarrow 0^+_1)$ strength with neutron number in the stable Kr isotopes. This was linked to the emergence of quadrupole deformation possibly fragmenting the strength. The data are shown in Fig.\,\ref{fig:systematics}\,(b). When inspecting Figs.\,\ref{fig:systematics}\,(a) and (b), it is apparent that the $B(E3;3^-_2 \rightarrow 0^+_1)$ follows the $B(E2;2^+_1 \rightarrow 0^+_1)$ strength increase. There is a caveat though. In the stable isotope \nuc{78}{Kr}, the $3^-_2$ is not the one reported in \cite{Mat82a}. There exists a lower-lying $3^-$ state at 2678\,keV \cite{ENSDF}, which was most likely not observed in $(p,p')$ because of its small $B(E3)$ strength [see Fig.\,\ref{fig:systematics}\,(c)]. The fragmentation of the $B(E3)$ strength appears, thus, nontrivial as quadrupole deformation begins to manifest. The observations made for \nuc{72}{Se} support this point. For the Kr isotopes, we will consequently use the notation $3^-_{2,(p,p')}$ for the second $3^-$ state observed in $(p,p')$ and carrying non-negligible $B(E3)$ strength. In unstable \nuc{76}{Kr}, the situation is comparable to \nuc{78}{Kr}. Matsuki {\it et al.} reported a possible $3^-$ state at 2601(15)\,keV \cite{Mat81a}, which they observed in the \nuc{78}{Kr}$(p,t)$\nuc{76}{Kr} reaction. In a follow-up publication, they presented a firmly identified $3^-$ state at 2872(15)\,keV as the $3^-_2$ candidate \cite{Mat82a}. We do not observe any resolved $\gamma$ rays in our spectra, which could be attributed to the population of the 2872-keV state in $(p,p')$. Its $B(E3)$ strength must either be small or its $\gamma$-decay behavior be very complex, i.e., the yield be shared between several $\gamma$-decay branches. Both scenarios may prevent its detection. A $J^{\pi} = 3^-$ assignment is possible for a state at 2581\,keV though. Previously, Giannatiempo {\it et al.} had argued for a $J^{\pi} = 2^+$ assignment based on the state's $\gamma$-decay behavior and the deduced $\log(ft)$ value \cite{Gia05a}. The $\log(ft) = 7.1$ value is, however, exactly the same as for the known $3^-_1$ state in \nuc{76}{Kr}. The corresponding feeding intensity is also low suggesting a forbidden decay from the $J^{\pi} = 1^-$ parent ground state of \nuc{76}{Rb}. Furthermore, the $\gamma$ decays to the $2^+_2$ ($I_{\gamma} = 78$\,$\%$) and $4^+_1$ ($I_{\gamma} = 22$\,$\%$) states allow for a $J^{\pi} = 3^-$ assignment. This state might, thus, correspond to the 2601-keV state reported by Matsuki {\it et al.}, where the $(p,t)$ angular distribution favors an $l = 3$ transfer. We observe the population of the 2581-keV state in $(p,p')$. The $(3^-_2) \rightarrow 2^+_2$ and $(3^-_2) \rightarrow 4^+_1$ transitions are highlighted in Fig.\,\ref{fig:spectra}. As mentioned earlier, the $\gamma$-decay intensities reported in Ref. \cite{Gia05a} were used for the \textsc{ucgretina} simulation. Even though the spin-parity assignment is tentative, the deduced $B(E3;3^-_{2,(p,p')} \rightarrow 0^+_1) = 6.2 \pm 0.7 \mathrm{(stat.)} \pm 0.7 \mathrm{(sys.)}$ W.u. fits well into the systematics. For \nuc{74}{Kr}, we observe a new $\gamma$-ray transition of 2062(5)\,keV (see Fig.\,\ref{fig:spectra}), which is in coincidence with the 456-keV, $2^+_1 \rightarrow 0^+_1$ transition. Thus, there is evidence for a previously unobserved level at 2518(5)\,keV. If this state is indeed a $3^-$ state, then this would establish two excited $3^-$ states within $\sim 100$\,keV. We, thus, want to emphasize that the first two excited $3^-$ states of the $N=38$ isotone \nuc{72}{Se} are within 28\,keV \cite{ENSDF}. The 2518-keV $\gamma$-ray yield corresponds to a $B(E3;3^-_{2,(p,p')} \rightarrow 0^+_1) = 4.3 \pm 1.1 \mathrm{(stat.)} \pm 0.5 \mathrm{(sys.)}$\,W.u., which is again in excellent agreement with the general trend seen in Fig.\,\ref{fig:systematics}\,(b). Interestingly, for both \nuc{74,76}{Kr} and even though tentatively assigned, the $B(E3;3^-_{2,(p,p')} \rightarrow 0^+_1)$ strength agrees with the upper limit of 4.5 W.u. determined for the $B(E3;3^-_1 \rightarrow 0^+_1)$ strength in \nuc{72}{Se}.

Based on our new data, we establish that the sudden $B(E3)$ strength increase at $N=40$ is exclusively observed for \nuc{72}{Ge} ($Z = 32$). For the $N=38$ isotones, it is observed for \nuc{70}{Ge} ($Z=32$) and \nuc{72}{Se} $(Z=34)$ but not for \nuc{74}{Kr} ($Z=36$). A simple picture of enhanced octupole correlations around the octupole magic number $N = 40$ can consequently not be claimed. The almost degenerate, low-lying $3^-$ states and the fact that -- in contrast to \nuc{70,72}{Ge}, \nuc{74}{Se} and \nuc{74,76}{Kr} -- the $3^-_2$ state is the strongest fragment in \nuc{72}{Se} also suggest that two microscopic configurations could cross beyond $A=74$.

 \begin{figure}[t]
\centering
\includegraphics[width=1\linewidth]{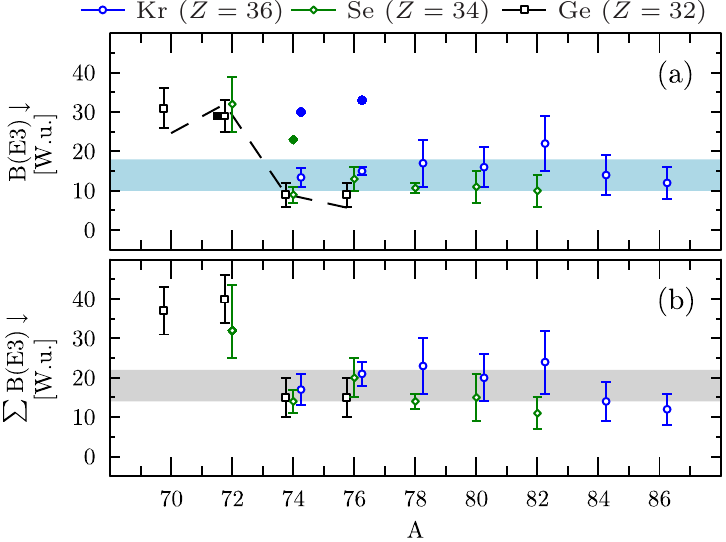}
\caption{\label{fig:summed}{(color online) Experimental (a) $B(E3;3^-_i \rightarrow 0^+_1)$ strength of the strongest fragment and (b) summed $\sum_i B(E3;3^-_i \rightarrow 0^+_1)$ strengths in Ge, Se and Kr isotopes up to 5\,MeV (open symbols with uncertainties). Except for \nuc{72}{Se} and \nuc{74,76}{Kr}, the data are from \cite{Kib02a, Ros86a, Sch87a, Ogi86a, Mat82a}. For all nuclei but \nuc{72}{Se}, the strongest fragment is the $3^-_1$ state. Theoretical predictions for the $B(E3;3^-_1 \rightarrow 0^+_1)$ strengths of \nuc{72}{Ge}, \nuc{74}{Se}, and \nuc{74,76}{Kr} were added to panel (a) [solid symbols]. These were obtained with the recently introduced configuration-mixing $sdf$ IBM mapping approach \cite{Nom22b}. The color coding and corresponding symbols are the same for the experimental data and theoretical predictions. Additionally, predictions made within IBM+IBFM approach of Ref. \cite{Chu93a} for the Ge isotopes were added to panel (a) [dashed line]. The blue and grey bands correspond to the 14(4) W.u. and 18(4) W.u. averages mentioned in the text.}} 
 \end{figure}

To obtain a clearer picture, the $B(E3;3^-_i \rightarrow 0^+_1)$ of the strongest fragment and summed $B(E3;3^-_i \rightarrow 0^+_1)$ strengths in the Ge-Kr mass region are compiled in Fig.\,\ref{fig:summed}. Except for \nuc{72}{Se} and \nuc{74,76}{Kr}, the summed strengths were deduced from the available proton and alpha inelastic scattering experiments on the stable Ge, Se and Kr isotopes \cite{Ros86a, Sch87a, Ogi86a, Mat82a}. The data draw an intriguing picture of two distinct regions. The first region extends from the spherical $N = 50$ neutron shell closure all the way down to $A = 74$. In this region, weighted averages of $B(E3) = 14(4)$ W.u. and $\sum B(E3) = 18(4)$ W.u. are determined from the data. The quoted uncertainties correspond to the standard deviation. The new data on \nuc{74,76}{Kr}, with their dominant prolate ground-state configuration \cite{Cle07a}, fit perfectly into this group. Then, the sudden jump of the $B(E3)$ strength is observed at $A=72$. The location of this ``jump'' coincides with the transition from a prolate to an oblate ground-state configuration at $A \approx 72$ \cite{Gad05a, Lju08a, Iwa14a, Wim20a, Hen18a}. However, based on experimental data, triaxial configurations appear to be important at $A = 72$, too \cite{Aya16a}. Most importantly, the ``jump'' is not observed at a fixed proton or neutron number as might be naively expected and, thus, seems to be more intimately connected to specific structure changes.

Predictions for the $B(E3;3^-_1 \rightarrow 0^+_1)$ strengths of \nuc{72}{Ge}, \nuc{74}{Se} and \nuc{74,76}{Kr}, obtained with the pioneering configuration-mixing $sdf$ IBM mapping approach \cite{Nom22b}, were added to Fig.\,\ref{fig:summed}\,(a). Before discussing these predictions, it should be mentioned that the experimentally determined magnitude of the quadrupole moments $Q_{2^+_1}$ are well reproduced while for all nuclei but \nuc{74}{Kr} the predicted signs disagree with the data. A similar observation was made in Ref. \cite{Hen18a}. As the self-consistent mean-field results already predict pronounced oblate minima, this gets propagated to the mapped IBM wave functions. However, Ref. \cite{Nom22b} also shows that the ground-state wave functions of all considered nuclei are strongly mixed with spherical 0p-0h, oblate 2p-2h and prolate 4p-4h configurations contributing. This highlights the complexity of this mass region. Still, for a more meaningful comparison in terms of octupole collectivity, a functional should be employed, which can reproduce the signs of the experimental quadrupole moments as the correct ground-state structure is critical \cite{Rob12a}. The consistent overprediction of the $B(E3)$ strengths, with the exception of \nuc{72}{Ge}, could consequently be an artifact of the incorrect and dominantly oblate ground-state structure. Most $3^-_1$ $sdf$-IBM wave functions are also predicted to be dominated by the oblate 2p-2h intruder configuration. More importantly though, the calculations show that enhanced octupole collectivity is indeed expected for the oblate configuration. This qualitatively agrees with the significant $B(E3)$ strength increase seen at $A=72$, where oblate configurations start to strongly mix into or even dominate the ground state wave function (see, {\it e.g.}, the work of Refs. \cite{Gad05a, Cle07a, Lju08a, Iwa14a, Hen18a, Wim20a, Wim21a, Hin10a, Sat11a, Rod14a}). For completeness, we added the IBM+IBFM results of Ref. \cite{Chu93a} to Fig.\,\ref{fig:summed}\,(a). As the parameters are, however, explicitly fitted to the Ge isotopes, no clear microscopic information for the entire Ge-Kr mass region can be extracted. Considering the projected shell model calculations of Ref.\,\cite{Wu17a}, which predict a dominant two-quasiparticle (2QP) character for the lowest negative-parity rotational bands in the Kr isotopes, it is possible that for nuclei with $A > 72$ the contribution of the $f_{5/2}-g_{9/2}$ fermion-pair configuration to the total wave function increases and leads to decreased octupole collectivity as in the Ge isotopes. The inspection of the predicted structures reveals, however, that the 2QP Nilsson configurations in \nuc{72-76}{Kr} originate from the spherical $2p_{3/2}$ and $1g_{9/2}$ orbitals, i.e., the octupole-collectivity driving orbitals \cite{Wu17a}. While the experimental signature is clear, the theoretical picture in the Ge-Kr mass region remains a puzzle.

\section{Summary}

In summary, we have performed inelastic proton scattering experiments in inverse kinematics on the rare isotopes \nuc{72}{Se} and \nuc{74,76}{Kr} to measure their $B(E3)$ strengths. While significantly enhanced octupole strength of $\sim 32$ W.u. was established for \nuc{72}{Se} ($Z = 34, N = 38$), much smaller strength of $\sim 15$ W.u. was observed for \nuc{74,76}{Kr} ($Z = 36, N = 38,40$). Based on our new data, we establish that the sudden $B(E3)$ strength increase at $N=40$ is exclusively observed for \nuc{72}{Ge} ($Z = 32$). For the $N=38$ isotones, it is observed for \nuc{70}{Ge} ($Z=32$) and \nuc{72}{Se} $(Z=34)$ but not for \nuc{74}{Kr} ($Z=36$). The almost degenerate, low-lying $3^-$ states and the fact that -- in contrast to \nuc{70,72}{Ge}, \nuc{74}{Se} and \nuc{74,76}{Kr} -- the $3^-_2$ state is the strongest fragment in \nuc{72}{Se} also suggest that two microscopic configurations could cross beyond $A=74$. In combination with previously existing data, the new data clearly question a simple origin of enhanced octupole strengths around $N = 40$. Instead, the present work establishes two regions of distinct octupole strengths with a sudden strength increase around the $A=72$ prolate-oblate-triaxial shape transitional point. Theoretical calculations performed in the framework of the configuration-mixing $sdf$ IBM mapping approach predict enhanced $B(E3)$ strengths built on the oblate minimum in this mass region, but fall short on correctly describing the ground-state structure of the considered nuclei. Future experiments at next-generation rare isotope beam facilities must test whether, as in \nuc{70}{Ge}, enhanced octupole strengths can also be observed in the $A = 70$ isobars \nuc{70}{Se} and \nuc{70}{Kr}. To investigate how far the region of enhanced octupole collectivity extends, strengths should also be determined for the even lighter Ge, Se and Kr isotopes. To arrive at a sound understanding of the experimental data, more microscopic calculations, along the lines of Ref. \cite{Nom22b} and which incorporate configuration mixing as well as triaxial degrees of freedom, are called for.

\begin{acknowledgments}
This work was supported by the National Science Foundation (NSF) under Grant No. PHY-2012522 (WoU-MMA: Studies of Nuclear Structure and Nuclear Astrophysics), Grant No. PHY-1565546 (NSCL), Grant No. PHY-2209429 (Windows on the Universe: Nuclear Astrophysics at FRIB), and by the Department of Energy, Office of Science, Office of Nuclear Physics, Grant No. DE-SC0020451 (MSU). GRETINA was funded by the Department of Energy, Office of Science. The operation of the array at NSCL was supported by the DOE under Grant No. DE-SC0019034. M.S. acknowledges support through the FRIB Visiting Scholar Program for Experimental Science 2020.
\end{acknowledgments}

\bibliography{Kr_02}

\end{document}